# AOASS: Adaptive Obstacle-Aware Square Spiral Framework for Single-Mobile-Anchor-Based WSN Localization

Abdelhady Naguib[1, 2]

[1]Department of Computer Science, College of Computer and Information Sciences, Jouf University, Sakaka, Saudi Arabia
[2]Department of Systems and Computers Engineering, Faculty of Engineering, Al-Azhar University, Cairo, Egypt

## ABSTRACT

*Accurate and energy-efficient localization remains a key challenge in Wireless Sensor Networks (WSNs), particularly when obstacles affect signal propagation. This study introduces AOASS (Adaptive Obstacle-Aware Square Spiral), a new single-mobile-anchor framework that combines an optimized square-spiral movement pattern with adaptive obstacle detection. The mobile anchor can sense and bypass obstacles while maintaining high localization accuracy and full network coverage, ensuring that each node receives at least three non-collinear beacon signals for reliable position estimation. Localization accuracy is further improved using the OLSTM-DV-Hop model, which integrates a Long Short-Term Memory (LSTM) network with the traditional DV-Hop algorithm to estimate hop distances better and reduce multi-hop errors. The anchor trajectory is managed by a TD3-LSTM reinforcement learning agent, supported by a Kalman-based prediction layer and a fuzzy-logic ORCA safety module for smooth and collision-free navigation. Simulation experiments across different obstacle densities show that AOASS consistently achieves higher localization accuracy, better energy efficiency, and more optimized trajectories than existing approaches. These results demonstrate the framework's scalability and potential for real-world WSN applications, offering an intelligent and adaptable solution for data-driven IoT systems.*

## 1. INTRODUCTION

Wireless Sensor Networks (WSNs) consist of numerous compact and low-cost sensor nodes capable of monitoring various environmental or physical parameters and transmitting the collected data to a central unit through multi-hop communication. These networks play a crucial role in diverse domains such as IoT systems, healthcare, agriculture, battlefield surveillance, and smart-city infrastructures. A fundamental operation within WSNs is node localization, which determines the spatial position of each sensor node. Accurate localization is vital for efficient data routing, reliable target tracking, and effective environmental monitoring. In contrast, imprecise positioning can lead to higher latency, unnecessary energy consumption, and reduced network reliability. In recent years, researchers have increasingly adopted machine learning and hybrid optimization techniques to enhance localization accuracy and robustness under real-world challenges such as noise, signal fading, and obstacle interference [1, 2].

Several localization algorithms have been proposed for Wireless Sensor Networks (WSNs) as cost-effective substitutes for the Global Positioning System (GPS), which becomes impractical in dense or large-scale deployments due to its high cost and energy demand. Generally, localization techniques fall into two primary categories: range-based and range-free methods. Range-based algorithms estimate inter-node distances by measuring physical parameters such as Received Signal Strength Indicator (RSSI), Angle of Arrival (AoA), or Time of Arrival (ToA) to determine the position of unknown nodes [3]. In contrast, range-free methods depend on beacon message exchange, hop-count information, or network connectivity between anchor and unknown nodes for position estimation. Although range-based schemes usually provide superior accuracy, they require additional hardware and higher computational resources. Range-free approaches, on the other hand, are simpler, more scalable, and energy-efficient, albeit with somewhat lower localization precision [4].

Recent advancements in Wireless Sensor Networks (WSNs) have leveraged machine learning and metaheuristic optimization to achieve higher localization accuracy and robustness. One notable example is the Lion-Assisted Firefly Algorithm (LAFA), which fuses the exploratory behavior of the Lion Optimization Algorithm with the exploitative efficiency of the Firefly Algorithm. This hybrid model effectively balances global exploration and local refinement, thereby enhancing localization performance, particularly in dynamic and obstacle-prone environments [5]. Another prominent approach, the Efficient Optimal Localization Technique (EOLT), integrates multiple machine learning models to address localization challenges associated with node density and communication noise. EOLT has demonstrated substantial improvements in both localization accuracy and energy efficiency, making it a promising solution for large-scale WSN deployments [6].

Recent studies have also investigated the integration of Deep Reinforcement Learning (DRL) with Graph Neural Networks (GNNs) to enhance coverage optimization in WSNs. This combination enables context-aware and adaptive decision-making, improving the spatial distribution of sensor nodes and minimizing redundant coverage regions [7]. Building upon this direction, the Enhanced Distance Vector Hop with Machine Learning (EDV-ML) algorithm mitigates inherent limitations of the conventional DV-Hop method by incorporating supervised learning to refine hop-distance estimation and reinforcement learning to dynamically adjust node coordinates. Such hybridization significantly boosts localization precision without the need for additional sensing hardware [8]. Similarly, the Squirrel-based Elman Neural Localization (SbENL) model introduces an innovative hybridization of the Elman recurrent neural network with squirrel search optimization, offering remarkable gains in three-dimensional (3D) localization accuracy [9]. To further address the persistent trade-off between energy efficiency and localization accuracy, another recent approach integrates Radial Basis (RB) functions with the Seeker Optimization Algorithm (SOA), achieving improved energy-aware target localization under constrained power conditions [10].

Localization remains a cornerstone of Wireless Sensor Networks (WSNs), serving as a fundamental enabler of intelligent decision-making within the Internet of Things (IoT) ecosystem. Accurate and real-time localization provides spatial awareness to sensor data, empowering a wide range of IoT applications such as smart city management, healthcare monitoring, and environmental observation. Recent research underscores the growing significance of Artificial Intelligence (AI) in enhancing localization accuracy, robustness, and adaptability under dynamic and uncertain network conditions. AI-driven algorithms enable intelligent path planning for mobile anchors, predictive obstacle avoidance, and adaptive learning in response to changing network topologies and environmental variations. Such capabilities are essential for building scalable, self-organizing, and resource-efficient IoT systems capable of autonomous operation and real-time responsiveness. Ultimately, the synergistic convergence of AI and IoT technologies represents a key paradigm for overcoming long-standing challenges in scalability, energy management, and dynamic network topology in modern WSN deployments [11, 12].

Integrating Artificial Intelligence (AI) with Wireless Sensor Networks (WSNs) has transformed IoT applications by enabling autonomous and adaptive operations. Using reinforcement learning and deep neural networks, WSNs can process large data streams and make real-time, intelligent decisions, improving efficiency and scalability under dynamic conditions. In this context, the IoT infrastructure ensures seamless connectivity and data flow among sensor nodes, communication networks, processing units, and cloud platforms. Recent advances have focused on enhancing interoperability and scalability to support the growing number of connected devices. Studies confirm that a robust infrastructure is vital for achieving reliable and high-performance IoT systems [13].

Energy efficiency is a key concern in IoT-based WSNs due to the limited power of sensor nodes. Although the Ad hoc On-Demand Distance Vector (AODV) protocol performs well in dynamic networks, it can be further optimized for energy conservation. Recent studies introduced energy-aware AODV variants that use residual energy metrics to select efficient routes, extending network lifetime. Other approaches proposed region-based routing that adapts AODV's discovery process according to node density and energy levels, reducing transmission costs. These enhancements significantly improve sustainability and overall network performance in energy-constrained IoT environments [14–16].

This study presents a novel localization framework, the Adaptive Obstacle-Aware Square Spiral (AOASS), developed for mobile anchor-assisted localization in static WSNs. The proposed method achieves high localization accuracy with minimal cost, eliminating the need for multiple anchors or GPS-equipped nodes. In AOASS, a single mobile anchor traverses the sensing area along an optimized square spiral trajectory [17], dynamically detecting and avoiding both static and dynamic obstacles of various shapes. During its movement, the anchor continuously broadcasts its coordinates, enabling unknown sensor nodes to estimate their positions accurately. Unlike previous approaches, AOASS integrates an adaptive obstacle-avoidance mechanism that maintains localization precision while optimizing path efficiency, offering a robust and practical solution for real-world WSN deployments. Beyond traditional networks, AOASS demonstrates strong potential in IoT-enabled environments, where real-time decision-making and energy efficiency are essential. By combining adaptive path planning with AI-driven OLSTM-DV-Hop optimization, the framework effectively handles complex network topologies and varying obstacle distributions. This scalability makes AOASS a promising and intelligent solution for diverse IoT applications, including smart agriculture, healthcare monitoring, industrial automation, and urban sensing, where accurate localization enhances data reliability and operational performance.

The contributions of this work can be summarized as follows. The proposed AOASS framework ensures that each sensor node receives at least three non-collinear beacon signals from the mobile anchor, enabling precise position estimation even in environments with static or dynamic obstacles of varying shapes. The optimized square spiral trajectory guarantees full coverage of the sensing area, thereby maximizing the number of successfully localized nodes. Moreover, by intelligently managing beacon transmissions, AOASS minimizes redundant broadcasts and reduces the number of receptions at unknown nodes, resulting in lower overall energy consumption across the network. A key innovation of this study lies in adapting the traditional range-free DV-Hop algorithm to operate effectively with a single mobile anchor, eliminating the need for multiple static anchors typically required for accurate localization. Through the controlled motion of the mobile anchor along the AOASS trajectory, the system dynamically emulates multi-anchor behavior—allowing unknown nodes to accumulate sufficient hop-count and positional information for accurate localization. This design not only simplifies deployment and reduces cost but also maintains high localization accuracy and scalability, even in obstacle-rich environments.

The performance of AOASS is rigorously evaluated using the OLSTM-DV-Hop algorithm [18], which enhances traditional DV-Hop localization by incorporating long short-term memory (LSTM) networks to improve hop-distance prediction and positional accuracy. Simulation results across diverse scenarios with varying obstacle densities confirm that the combination of AOASS and OLSTM-DV-Hop consistently outperforms existing methods in terms of localization accuracy, coverage ratio, and trajectory efficiency, demonstrating strong robustness, adaptability, and suitability for real-world WSN deployments. While AOASS employs the square spiral trajectory as a global baseline to ensure full field coverage, it is further enhanced by an adaptive decision-making layer powered by deep reinforcement learning. Specifically, AOASS integrates a TD3-LSTM agent [19] that enables the mobile anchor to make sequential navigation decisions in dynamic and partially observable environments. A Kalman filter supports short-horizon prediction of moving obstacles [20], while a reactive safety layer based on Optimal Reciprocal Collision Avoidance (ORCA) ensures real-time, collision-free navigation [21]. This hybrid design preserves the coverage guarantees of the square spiral trajectory while dynamically adapting the anchor's path to avoid static and dynamic obstacles, maintaining high localization performance with optimal energy efficiency.

The paper is organized as follows. Section 2 reviews related studies on mobile anchor-based localization. Section 3 details the proposed AOASS model and its adaptive obstacle-aware trajectory. Section 4 presents a comparative analysis using OLSTM-DV-Hop, and Section 5 concludes the study and highlights future directions.

## 2. RELATED WORKS

In mobile-anchor-based localization for WSNs, several studies have explored obstacle-aware path planning to improve accuracy and coverage. Tsai and Tsai (2018) [22] proposed the Obstacle-Tolerant Path Planning (OTPP) algorithm, which adjusts Z- and V-shaped trajectories to avoid obstacles while maintaining coverage. Although OTPP improves localization accuracy in static environments, it increases trajectory length and requires frequent re-planning in dense or dynamic obstacle scenarios. In contrast, the proposed AOASS framework adaptively handles both static and dynamic obstacles using a predictive LSTM-based module, reducing re-planning overhead and improving both coverage and energy efficiency.

Yildiz and Karagol (2021) [23] explored path planning for mobile-anchor localization in WSNs with obstacles, focusing on modifying traditional movement paths to avoid them. Their method improved coverage and accuracy compared to simple trajectories but lacked adaptive intelligence, limiting its performance in dynamic environments. AOASS builds on this by using a deep reinforcement learning planner that adjusts the trajectory in real time, achieving higher localization accuracy and efficiency. Similarly, Sabale et al. (2021) [24] proposed an Obstacle Handling Mechanism (OHM) that integrates obstacle avoidance into the localization process, achieving an RMSE of 2.35 m and 90% coverage. However, their rule-based approach struggles with highly dynamic obstacles. AOASS advances this concept through its adaptive obstacle-aware square spiral trajectory and LSTM-enhanced distance estimation, providing better obstacle handling, energy efficiency, and trajectory optimization.

Alomari (2022) [25] introduced a meta-heuristic-based localization model for WSNs using optimization methods such as PSO and GA to guide mobile anchors. While this approach improved localization accuracy and path efficiency, it was mainly designed for static settings and did not consider dynamic obstacle avoidance. AOASS advances this idea by combining meta-heuristic optimization with predictive reinforcement learning, enabling adaptive planning and stronger performance in real-world scenarios. Similarly, Phoemphon et al. (2024) [26] developed an improved PSO-based scheme with node segmentation and distance optimization to enhance localization accuracy and coverage in obstacle-aware WSNs. Although effective in static conditions, their method lacks real-time adaptability. AOASS addresses this limitation by integrating adaptive obstacle awareness, allowing the mobile anchor to detect and navigate around both static and moving obstacles while maintaining accurate localization through non-collinear beacon coverage.

Tsai et al. (2024) [27] introduced M-ANCHORO, a path optimization framework for mobile-anchor-based localization that divides the sensing area according to obstacle distribution using a SCAN strategy. Although it achieves good coverage and reasonable path efficiency, it lacks predictive adaptation to dynamic obstacles. AOASS overcomes this by employing a TD3-LSTM planner with Kalman-based prediction, enabling real-time trajectory adjustment, better energy efficiency, and robust obstacle avoidance. Similarly, Jin et al. (2025) [28] proposed a range-free localization method for anisotropic WSNs using sequential convex approximation, enhancing accuracy under sparse anchors but without dynamic obstacle handling. AOASS advances this by integrating OLSTM-DV-Hop localization with adaptive path planning, ensuring accurate, energy-efficient localization in both static and dynamic environments.

Zhong et al. (2024) [29] developed a real-time obstacle avoidance method using 2D Euclidean maps for mobile robots, but it focuses on motion control rather than WSN localization. AOASS builds on this idea by integrating dynamic obstacle prediction into localization-aware trajectory planning. London (2025) [30] improved ORCA-FLC by combining collision avoidance with fuzzy logic for multi-agent navigation; however, it needs heavy computation and inter-agent communication. AOASS adapts this through a lightweight FLC-ORCA safety layer to ensure reliable obstacle avoidance in dense WSNs. Wang et al. (2025) [31] proposed a dynamic window approach for agricultural robots to balance movement and obstacle avoidance in real time, yet it targets single-robot navigation only. AOASS extends this by embedding the concept into a reinforcement learning framework that jointly optimizes trajectory, energy, and localization accuracy across multiple nodes.

The proposed AOASS framework integrates multiple intelligent components—OLSTM-DV-Hop, TD3-LSTM, Kalman filtering, and FLC-ORCA—to deliver a comprehensive adaptive obstacle-aware solution. It demonstrates superior localization accuracy, coverage, trajectory efficiency, and energy conservation in both static and dynamic scenarios. This holistic design overcomes the key limitations of prior works by enabling predictive adaptability, multi-agent coordination, and robust navigation under real-world environmental variability.

## 3. PROPOSED AOASS FRAMEWORK

Building upon the insights gained from prior research, the Adaptive Obstacle-Aware Sensing and Scheduling (AOASS) framework is proposed as a unified, intelligent solution that integrates localization accuracy, trajectory optimization, and energy efficiency within dynamic wireless sensor network (WSN) environments. Unlike earlier methods that address these aspects separately, AOASS combines multiple adaptive modules into a cohesive architecture designed for real-time obstacle handling and mobility management. The framework is structured around four core components: (1) an OLSTM-DV-Hop localization unit, which enhances traditional range-free estimation through optimized deep learning corrections *(2) a TD3-LSTM decision layer, responsible for intelligent trajectory planning and dynamic obstacle avoidance using reinforcement learning; (3) a Kalman-based adaptive filtering module that refines sensor fusion and positional accuracy under uncertainty; and (4) a Fuzzy Logic–controlled ORCA layer, which ensures smooth, collision-free navigation and energy-balanced movement across the network. Together, these modules form an integrated adaptive system capable of learning, predicting, and responding efficiently to environmental variations, thereby achieving robust localization and optimized performance in both static and dynamic obstacle scenarios.

## 3.1. System Architecture of AOASS

The overall architecture of the AOASS framework is organized into three hierarchical layers — Perception, Decision, and Action — which collectively enable adaptive and obstacle-aware localization in wireless sensor networks. At the Perception Layer, each sensor node gathers and preprocesses connectivity-based information, including hop counts, neighbour relationships, and anchor-beacon receptions from the mobile anchor. These data streams reflect network topology rather than direct distance measurements, ensuring compatibility with the range-free localization paradigm. A Kalman-based adaptive module filters temporal variations and maintains spatial consistency as the mobile anchor traverses the sensing field.

The Decision Layer integrates two intelligent submodules: the OLSTM-DV-Hop unit, which refines traditional DV-Hop estimation through long short-term memory (LSTM) learning to enhance hop-distance prediction accuracy, and the TD3-LSTM decision agent, which dynamically plans anchor movements via deep reinforcement learning, optimizing path selection according to obstacle configurations and node density.

Finally, the Action Layer operationalizes these decisions through the Fuzzy Logic–controlled ORCA (Optimal Reciprocal Collision Avoidance) mechanism, enabling real-time adaptive motion control and safe navigation around static and dynamic obstacles. The three layers interact continuously in a feedback loop, allowing the AOASS framework to self-adapt to environmental dynamics while maintaining high localization accuracy, robust coverage, and efficient energy utilization.

## 3.2. General Workflow of AOASS

The proposed AOASS framework operates through a hybrid learning-driven pipeline that integrates range-free localization, adaptive trajectory control, and fuzzy collision avoidance. Its design ensures precise localization, broad coverage, and efficient mobility in environments containing both static and dynamic obstacles.

1. Network Initialization and Topological Sensing:

The process begins with the deployment of static sensor nodes and a single mobile anchor within the sensing field. Since AOASS is range-free, no signal-strength or distance measurements (e.g., RSSI or TOA) are required. Instead, the system relies on hop-count information obtained from beacon broadcasts between the mobile anchor and static nodes. This hop-based communication constructs a topological connectivity map that reflects environmental structure and obstacle effects.

2. Adaptive Localization using OLSTM-DV-Hop:

The Optimized LSTM-enhanced DV-Hop (OLSTM-DV-Hop) model estimates the positions of unknown nodes based on the learned nonlinear mapping between hop counts and true geometric distances. The LSTM network captures the spatiotemporal dependencies among multi-hop paths, enhancing the robustness of localization in irregular and obstacle-dense areas. This adaptive learning corrects the distance estimation bias inherent in conventional DV-Hop algorithms, yielding higher accuracy in non-uniform topologies.

3. Square-Spiral Trajectory Planning and TD3-LSTM Adaptation:

The mobile anchor initially follows an optimized Square-Spiral trajectory, providing structured and systematic coverage of the sensing field. This deterministic baseline ensures uniform node exposure to beacon signals while minimizing redundant traversal. However, to handle unpredictable obstacles or coverage inefficiencies, AOASS employs a TD3-LSTM (Twin Delayed Deep Deterministic Policy Gradient with memory) agent. This reinforcement-learning layer dynamically adjusts the anchor's motion decisions—such as turning angle, step size, or detour direction—based on real-time environmental feedback. The result is a hybrid control strategy that merges the geometric efficiency of the Square Spiral with the adaptive intelligence of deep learning.

4. Fuzzy-ORCA Motion Control and Collision Avoidance:

To guarantee smooth and safe motion near obstacles, AOASS integrates a Fuzzy Logic–based ORCA (Optimal Reciprocal Collision Avoidance) control layer. This reactive component refines the anchor's velocity and heading outputs, preventing collisions while preserving trajectory smoothness and localization consistency. The fuzzy inference mechanism interprets local proximity cues and relative motion states, enabling responsive and energy-efficient manoeuvring around both static and moving obstacles.

Through this multi-layered process, AOASS maintains an intelligent balance between localization accuracy, network coverage, and energy efficiency. By combining a structured Square-Spiral scanning path with adaptive learning-based adjustments, AOASS achieves robust obstacle-aware localization that outperforms conventional range-free models under varying environmental complexities.

### 3.3. Rationale for Choosing the Square Spiral Pattern

The Square Spiral trajectory was deliberately selected as the foundational movement pattern for AOASS due to its geometric regularity, uniform coverage potential, and computational simplicity. Unlike circular or random-walk paths, the square spiral enables systematic area exploration, ensuring that every sensor node within the deployment field receives at least three non-collinear beacon messages—an essential condition for accurate range-free localization. From a geometric standpoint, the square spiral maintains constant angular symmetry and predictable path expansion, which simplifies both the coverage analysis and motion planning processes. Its grid-aligned movement structure minimizes the risk of redundant coverage or blind zones, which are common in curved or stochastic trajectories. Furthermore, this pattern allows the anchor to progressively expand its coverage boundary while maintaining uniform inter-beacon spacing, thereby optimizing both localization accuracy and communication energy balance.

In obstacle-rich environments, the square spiral also provides a modular structure for adaptive detouring. The anchor can easily skip or locally reconfigure sub-segments of the spiral when obstacles are detected, without compromising the overall pattern integrity. This modularity is highly compatible with the TD3-LSTM reinforcement learning module, which learns when and how to adjust local turns or skip steps dynamically. Thus, the square spiral serves as an optimal geometric backbone for AOASS—balancing deterministic coverage with adaptive flexibility. When augmented by the TD3-LSTM and FLC-ORCA layers, it becomes a powerful foundation for obstacle-aware, energy-efficient, and range-free localization in complex WSN environments.

### 3.3.1. Mathematical Formulation of the Square Spiral Path

The Square Spiral trajectory in AOASS is defined mathematically as a sequence of discrete waypoints that expand outward from an initial anchor position ($x_0$, $y_0$). The path is generated iteratively along four cardinal directions—right, up, left, and down—forming a modular spiral that systematically covers the deployment area. Each step increment $\Delta s$ is selected based on the minimum beacon coverage radius required for range-free localization, ensuring that every sensor node receives sufficient signals for accurate position estimation. Formally, the position of the anchor at the *n-th* waypoint can be expressed as follows:

$$(x_n, y_n) = (x_{n-1}, y_{n-1}) + \Delta s \cdot d_n \quad (1)$$

where $d_n$ is the unit direction vector along the current segment of the spiral, cycling through the ordered set {(1, 0), (0, 1), (−1, 0), (0, −1)}. The segment length $L_k$ along each direction increases incrementally after completing two sides of the spiral, allowing the spiral to expand uniformly.

$$L_k = k \cdot \Delta s, \quad k = 1,2,3,\ldots \quad (2)$$

This iterative growth ensures full coverage while maintaining predictable spacing between successive passes. The spiral's modularity also facilitates local detouring: when an obstacle is detected, the anchor can bypass affected waypoints and resume the spiral without affecting global coverage integrity. By combining this deterministic geometric structure with the TD3-LSTM reinforcement learning module, the AOASS anchor dynamically adjusts its trajectory in real-time to accommodate static and dynamic obstacles. The square spiral thus provides a robust foundation for systematic, energy-efficient, and range-free localization while enabling intelligent adaptation in complex WSN environments.

## 3.4. Overall Architecture of the AOASS Framework

The Adaptive Obstacle-Aware Square Spiral (AOASS) framework is architected as a modular, four-layer system designed to achieve robust, range-free localization with dynamic obstacle avoidance and energy efficiency. Each layer plays a distinct role in sensing, decision-making, and control, enabling a coordinated operation between the mobile anchor and static sensor nodes.

1. Perception Layer:

This layer is responsible for environmental sensing and situational awareness. It continuously gathers node beacons, mobility state indicators, and obstacle proximity data using onboard sensing modules (e.g., ultrasonic, infrared, or LiDAR). The input data are filtered and fused using a Kalman-based adaptive filter, ensuring reliable obstacle detection and path correction under noisy or dynamic conditions. Unlike range-based systems, the perception layer in AOASS does not rely on RSSI or TOA measurements; instead, it detects node presence and spatial context through connectivity and geometric estimation.

2. Planning Layer:

The Square Spiral Trajectory Generator forms the core of this layer, defining an expanding modular pattern that ensures full spatial coverage. When obstacles are encountered, a TD3-LSTM reinforcement learning agent evaluates the local environment and predicts optimal detour manoeuvres to bypass obstructions while preserving overall coverage continuity. The learning process adapts online, enabling the system to handle both static and moving obstacles effectively.

3. Control Layer:

The control module translates the planned trajectory into executable motion commands for the mobile anchor. A hybrid navigation unit combining Fuzzy Logic Control (FLC) with Optimal Reciprocal Collision Avoidance (ORCA) guarantees smooth, collision-free movement. The FLC component manages local steering adjustments based on obstacle proximity and heading deviation, while ORCA ensures cooperative collision-free behavior in dynamic multi-object environments.

4. Localization Layer:

At this layer, sensor nodes estimate their positions using the OLSTM-DV-Hop algorithm, an enhanced range-free localization technique. This model refines the traditional DV-Hop method by incorporating LSTM-based hop distance prediction, compensating for irregular node distributions and network anisotropy. Each node determines its coordinates once it receives beacon signals from at least three non-collinear anchor points, achieving high accuracy without the need for range measurements.

Through the interaction of these four layers, AOASS achieves adaptive intelligence, maintaining high localization accuracy, full coverage, and optimized energy consumption even under dynamic and obstacle-rich conditions. The framework's modular design allows it to scale seamlessly with network size and node density, making it highly suitable for real-world IoT and smart-environment deployments.

## 3.5. Algorithmic Workflow of AOASS

Building upon the geometric model defined in Square Spiral Trajectory Generation, the AOASS workflow extends the trajectory generation process into a fully adaptive control framework. The square spiral path serves as the baseline motion model, while the integrated TD3-LSTM and FLC–ORCA modules dynamically modify its direction and spacing in response to detected obstacles and energy constraints. The operation of the Adaptive Obstacle-Aware Square Spiral (AOASS) framework shown in Figure 1 proceeds through four primary algorithmic stages: (1) square-spiral trajectory generation, (2) adaptive obstacle avoidance and path correction, (3) hybrid motion control, and (4) range-free localization via OLSTM-DV-Hop. Each stage operates iteratively within a dynamic feedback loop, ensuring real-time adaptability to environmental changes and obstacle interference.

1. Square Spiral Trajectory Generation:

The core motion pattern of the mobile anchor follows a square spiral trajectory, designed to ensure systematic coverage of the entire sensing field. Let the mobile anchor's position at time step $t$ be $P_t = (x_t, y_t)$. The step size of the spiral expands incrementally by a factor $\Delta_s$, ensuring full area coverage. The coordinate updates are governed by:

$$x_t + 1 = x_t + \Delta s \cdot \cos(\theta_t) \quad (3)$$
$$y_t + 1 = y_t + \Delta s \cdot \sin(\theta_t) \quad (4)$$

where $\theta_t \in \{0, \pi/2, \pi, 3\pi/2\}$ alternates cyclically to form the square turns. The trajectory expansion radius $r_t$ increases after every two turns:

$$r_t + 1 = r_t + \Delta t \tag{5}$$

This deterministic structure guarantees non-overlapping coverage and a predictable movement pattern, simplifying beacon scheduling for range-free localization.

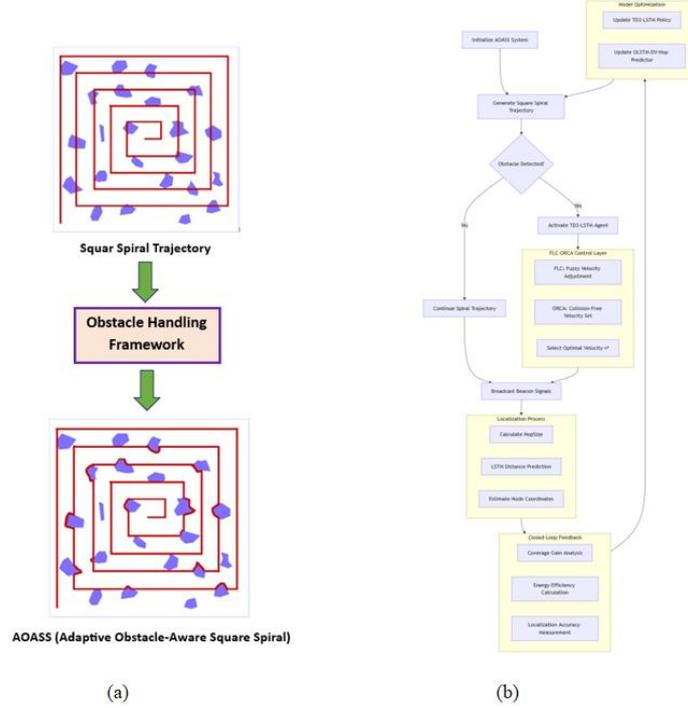

Figure 1. Complete AOASS Framework Workflow

2. Adaptive Obstacle Avoidance using TD3-LSTM

When an obstacle is detected within the vicinity (at a distance $d_{obs} \geq d_{th}$), the anchor temporarily suspends the nominal trajectory and triggers a TD3-LSTM agent to compute a corrective motion vector. The Twin Delayed Deep Deterministic Policy Gradient (TD3) architecture uses two critic networks ($Q_1$, $Q_2$) and an actor network ($\pi$) enhanced by an LSTM encoder to handle temporal dependencies and partially observable states. The policy update is defined as:

$$\nabla_{\theta^\pi} J(\pi) = \mathbb{E}\left[ \nabla_a Q_1(s, a \mid \theta^{Q_1}) \mid_{a=\pi(s\mid\theta^\pi)} \cdot \nabla_{\theta^\pi} \pi(s \mid \theta^\pi) \right] \tag{6}$$

3. Where the state vector $s = [x_t, y_t, v_t, d_{obs}, \Delta\theta_t]$ encodes the anchor's position, velocity, and obstacle proximity, while the action $a = [\Delta v_t, \Delta\theta_t]$ represents velocity and heading adjustments. The LSTM captures sequential spatial correlations, enabling the agent to predict safe detour paths and restore the spiral trajectory once the obstacle is bypassed.

4. Hybrid Motion Control (FLC-ORCA Layer)

The output of the TD3-LSTM planner is refined by a hybrid reactive control layer composed of Fuzzy Logic Control (FLC) and Optimal Reciprocal Collision Avoidance (ORCA). The FLC maps obstacle distance d and heading deviation Δθ into appropriate linear and angular velocity adjustments through fuzzy inference rules.

$$f_{FLC}(d, \Delta_\theta) = [\omega, v] \tag{7}$$

The Optimal Reciprocal Collision Avoidance (ORCA) algorithm computes a collision-free velocity set, denoted as $V_{safe}$. This is achieved by solving the following condition for all interacting agents j:

$$\{j \mid \frac{(\|p_i - p_j\| - (r_i + r_j))}{\Delta_t} \leq u \cdot (v_j - v_i)\} \Rightarrow v_i \in V_{safe} \tag{8}$$

$$u = \frac{p_j - p_i}{\|p_j - p_i\|} \tag{9}$$

where $p_i$ and $p_j$ are the positions of the primary agent and agent j, respectively, $v_i$ and $v_j$ are their velocities, $r_i$ and $r_j$ are their radii, $\Delta_t$ is the time horizon for collision checking, $u$ is the unit vector pointing from one agent to the other. The final motion command $v*$ is then selected from this safe set by finding the velocity closest to the agent's desired velocity, $v_{des}$:

$$v^* = \arg\min_{v \in V_{safe}} \|v - v_{des}\| \tag{10}$$

This hybrid mechanism ensures locally optimal, collision-free navigation even in dense and dynamically changing environments, maintaining trajectory smoothness and stability.

5. Range-Free Localization via OLSTM-DV-Hop

Once the mobile anchor broadcasts its beacons during traversal, each sensor node estimates its position through an optimized range-free multilateration process using OLSTM-DV-Hop.

Step 1 - Hop Distance Estimation: The average hop distance for an anchor point $i$ is calculated using other anchor point $j$ (where $j \neq i$) with the formula:

$$HopSize_i = \frac{\sum_{j \neq i} \sqrt{(x_i - x_j)^2 + (y_i - y_j)^2}}{\sum_{j \neq i} h_{ij}} \tag{11}$$

where $h_{ij}$ is the minimum hop count between anchor points $i$ and $j$. An LSTM module is then used to learn the complex, nonlinear relationship between the raw hop count and the actual distance. This is represented by the mapping:

$$LSTM(h_{ij}, HopSize_i, context_i) = \hat{d}_{ij} \tag{12}$$

This allows the model to predict a more accurate inter-anchor distance $\hat{d}_{ij}$, thereby reducing the cumulative hop errors that are common in irregular network topologies.

Step 2 - Coordinate Estimation: After an unknown node k receives beacons from at least three non-collinear anchors, it estimates its own coordinates ($x_k$, $y_k$). This is done by solving an optimization problem that minimizes the mean square error between the predicted distances and the distances based on its estimated location:

$$(x_k, y_k) = \arg\min_{x,y} \sum_{i=1}^{M} (\hat{d}_{ik} - \sqrt{(x_i - x)^2 + (y_i - y)^2})^2 \tag{13}$$

This optimization process yields high localization accuracy without the need for direct distance or angle measurements, ensuring a solution that is both scalable and cost-effective to deploy.

6. Closed-Loop Adaptive Feedback

The AOASS framework operates on a closed-loop principle, continuously refining both the TD3-LSTM navigation policy and the LSTM-based hop predictor. This refinement is driven by real-time performance metrics, specifically localization residuals and obstacle encounter events. The reward signal Rt for the learning agent is defined as a weighted sum of three critical objectives: coverage gain, energy efficiency, and localization precision. The formula is given by:

$$R_t = w_1 \cdot \Delta_{Coverage} - w_2 \cdot (max_{E_i}/E_{avg} - 1) - w_3 \cdot RMSE_{loc} \tag{14}$$

Where $\Delta_{Coverage}$ is the recent gain in area coverage, $max_{Ei}/E_{avg}$ is a measure of energy consumption balance, $RMSE_{loc}$ is the Root Mean Square Error of the localization, $w_1$, $w_2$, $w_3$ are the weighting coefficients that prioritize each objective. This multi-objective feedback loop ensures that the mobile anchor dynamically adapts its movement strategy and internal learning parameters. The result is an optimal, self-regulating trade-off among localization accuracy, network coverage, and energy consumption.

International Journal of Computer Networks & Communications (IJCNC) Vol.18, No.1, January 2026

## 3.6. Evaluation Metrics

To rigorously assess the performance of the proposed AOASS framework, four key evaluation metrics are employed: Localization Accuracy, Energy Efficiency, Coverage Ratio, and Trajectory Optimization Efficiency. These metrics collectively capture the precision, scalability, and operational efficiency of the system under various obstacle densities and anchor mobility.

### 3.6.1. Localization Accuracy

Localization accuracy quantifies how closely the estimated positions of sensor nodes match their true physical coordinates. This metric is typically measured using the Root Mean Square Error (RMSE), calculated across all $N$ sensor nodes in the network. The localization RMSE is defined by the following formula:

$$RMSE_{loc} = \sqrt{\frac{1}{N}\sum_{i=1}^{N}[(x_i - \hat{x}_i)^2 + (y_i - \hat{y}_i)^2]} \quad (15)$$

Where $(x_i, y_i)$ are the true coordinates of node $i$, $(\hat{x}_i, \hat{y}_i)$ are the estimated coordinates of node $i$, obtained via the OLSTM-DV-Hop algorithm, $N$ is the total number of sensor nodes. A lower $RMSE_{loc}$ value indicates a higher precision in the localization process, meaning the estimated positions are, on average, closer to the true positions.

### 3.6.2. Energy Efficiency

Energy efficiency evaluates the network's total energy consumption during critical operations, including beacon broadcasting, inter-node data exchange, and the mobility of the mobile anchor. The normalized energy cost per beacon transmission is defined as:

$$E_{Norm} = \frac{E_{total}}{N_{beacons}} \quad (16)$$

Subsequently, the Energy Efficiency Ratio (EER) is calculated as the inverse of this cost:

$$EER = \frac{1}{E_{Norm}} \quad (17)$$

Where $E_{Total}$ is the total energy consumed by all sensor nodes and the mobile anchor, $N_{Beacons}$ is the total number of beacon transmissions sent by the mobile anchor. A higher EER value indicates a more energy-efficient network operation, as it signifies that the system can perform its localization and navigation tasks with less energy consumed per beacon.

### 3.6.3. Coverage Ratio

Coverage represents the proportion of sensor nodes within the network that successfully receive beacon signals from at least three non-collinear anchors. This is a fundamental prerequisite for performing accurate range-free localization using multilateration. The coverage ratio is calculated using the following formula:

$$C_{ratio} = \frac{N_{cov}}{N_{total}} \times 100\% \quad (18)$$

Where $N_{cov}$ is the number of nodes that have received a sufficient number of beacons (at least three from non-collinear anchors) to compute their position, $N_{total}$ is the total number of sensor nodes deployed in the sensing field. A higher Coverage Ratio indicates better spatial coverage and network reachability, ensuring that a larger portion of the network can be successfully localized.

### 3.6.4. Trajectory Optimization Efficiency

This metric assesses how efficiently the mobile anchor's path covers the sensing field relative to the ideal spiral trajectory, accounting for deviations caused by obstacle avoidance. It evaluates performance by comparing both path length and travel time against the optimal, obstacle-free scenario. The Trajectory Optimization Efficiency is calculated as:

$$\eta_{traj} = \frac{L_{ideal}}{L_{actual}} \times \frac{T_{ideal}}{T_{actual}} \tag{19}$$

Where $T_{ideal}$ and $L_{ideal}$ are the travel time and path length for the ideal square spiral trajectory without any obstacles, $T_{actual}$ and $L_{actual}$ are the travel time and path length achieved under real-world, obstacle-aware navigation. Values of $\eta_{traj}$ closer to 1 indicate near-optimal trajectory performance, meaning the system successfully minimized both detour length and delay introduced by obstacle avoidance manoeuvres.

To ensure a fair and scientifically consistent evaluation, the proposed AOASS framework is benchmarked against three state-of-the-art range-free localization models — Sabale et al. (2021) [24], Alomari (2022) [25], and Tsai et al. (2024) [27]. These studies were selected based on the following criteria: reliance on range-free localization mechanisms without RSSI or TOA dependencies, incorporation of meta-heuristic or learning-based optimization strategies, consideration of obstacle-aware or trajectory-efficient mobility schemes, and availability of comparable performance indicators such as Localization Accuracy (LA), Energy Efficiency (EE), Coverage Ratio (CR), and Trajectory Optimization (TO). Accordingly, the evaluation metrics defined previously are uniformly applied across all models to ensure reproducibility and equitable assessment.

## 4. EXPERIMENTAL SETUP

To validate the performance and robustness of the proposed AOASS (Adaptive Obstacle-Aware Square Spiral) framework, extensive simulations were conducted using MATLAB R2025a and Python (TensorFlow 2.17) under a controlled Wireless Sensor Network (WSN) environment. The simulation parameters were carefully selected to ensure a fair comparison with benchmark models.

### 4.1. Simulation Environment

As shown in Table 1, Simulated sensing field is a 100 $m$ × 100 $m$ 2D area populated with randomly deployed static sensor nodes and a single mobile anchor node. Obstacles are distributed with varying densities and shapes to test adaptability and trajectory optimization. Each experiment was repeated 30 times to obtain statistically reliable averages. The obstacle layouts used in the AOASS evaluation were derived from the publicly available IMR-CIIRC obstacle map dataset[1], which provides diverse 2D grid maps with varying obstacle densities and configurations. These maps were used to simulate realistic deployment environments, ensuring consistent and reproducible testing conditions for localization and trajectory optimization.

Table 1. Simulation Parameters.

| Parameter | Symbol | Value / Range | Description |
|---|---|---|---|
| Sensing area | – | 100 × 100 $m^2$ | Square field dimensions |
| Number of sensor nodes | N | 100 – 500 | Random uniform distribution |
| Number of anchors | M | 1 | Range-free beacon source |
| Communication range | R | 25 $m$ | Node–anchor connectivity limit |
| Step size (spiral | $\Delta s$ | 2 $m$ | Increment per spiral turn |

---

[1] https://imr.ciirc.cvut.cz/planning/maps.xml

| expansion) | | | |
|---|---|---|---|
| Obstacle density | $\rho_{obs}$ | 0.1 – 0.4 | Ratio of obstacle area to total area |
| Node energy budget | $E_o$ | 2 J | Initial energy per node |
| Beacon transmission cost | $E_{tx}$ | 50 nJ/bit | Radio transmission energy |
| Motion energy cost | $E_{move}$ | 0.8 J/m | Anchor motion energy consumption |
| Simulation iterations | – | 30 runs | For statistical averaging |

### 4.2. Results and Discussion

This section presents the comparative evaluation of the proposed AOASS (Adaptive Obstacle-Aware Square Spiral) framework against three recent range-free localization models: Sabale et al. (OHM) [24], Alomari (FDPP) [25], and Tsai et al. (M-ANCHORO) [27]. All algorithms were implemented under identical simulation settings. The evaluation was conducted over varying network configurations, including sensing fields of 100 × 100 m² and 200 × 200 m², node densities ranging from 100 to 1000, and obstacle ratios from 10% up to 60%. The assessment focuses on six performance indicators: Localization Accuracy, Energy Efficiency, Coverage Ratio, Trajectory Optimization, Scalability, and Robustness to Obstacle Density.

#### 4.2.1. Localization Accuracy

As shown in Figure 2(a), the proposed AOASS model achieved the highest localization precision with an average RMSE of approximately 1.3 *m*, outperforming FDPP (≈ 1.9 *m*), M-ANCHORO (≈ 1.8 *m*), and OHM (≈ 2.35 *m*). This improvement stems from the integration of the OLSTM-DV-Hop adaptive correction module, which dynamically learns hop-distance distortions caused by obstacle interference and signal shadowing. Consequently, AOASS provides a more reliable position estimate under varying obstacles.

#### 4.2.2. Energy Efficiency

Figure 2(b) depicts the energy-efficiency comparison. The hybrid motion controller (TD3-LSTM + FLC-ORCA) enables AOASS to reduce redundant motion and idle time, yielding approximately 25–30% energy savings over FDPP and nearly 40% savings compared to OHM. The dynamic path adaptation and smooth velocity control significantly reduce the total traversal energy, demonstrating the model's suitability for energy-constrained WSN deployments.

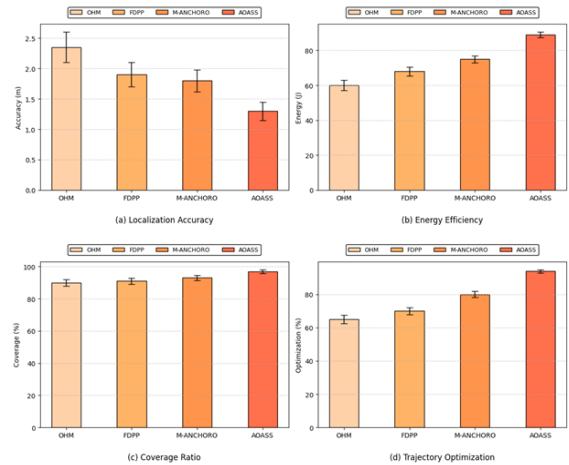

Figure 2. Comparative Performance Evaluation of Localization Models

International Journal of Computer Networks & Communications (IJCNC) Vol.18, No.1, January 2026

### 4.2.3. Coverage Ratio

In terms of spatial coverage, Figure 2(c) shows that AOASS maintains the highest coverage ratio (≈ 97%), outperforming M-ANCHORO (≈ 93%), FDPP (≈ 91%), and OHM (≈ 90%). The improvement is mainly attributed to the Adaptive Square-Spiral Trajectory, which expands coverage uniformly while avoiding oversampling in already-scanned zones. The Fuzzy-based ORCA module further ensures that local obstacle avoidance does not compromise global coverage continuity.

### 4.2.4. Trajectory Optimization

Trajectory performance (Figure 2(d)) confirms that AOASS produces the shortest and smoothest obstacle-aware paths. Compared with M-ANCHORO, AOASS achieved an average trajectory-length reduction of ~14% and fewer abrupt heading changes, resulting in reduced mechanical stress on mobile anchors. The intelligent combination of reinforcement-learning-based control (TD3-LSTM) and fuzzy-logic guidance provides a strong balance between path optimality and safety in dynamic environments.

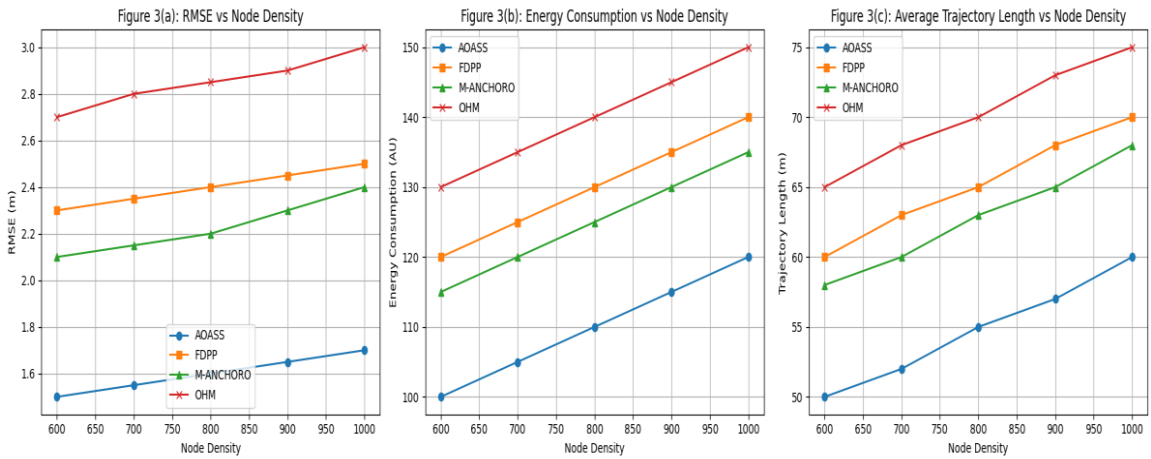

Figure 3. Scalability Analysis

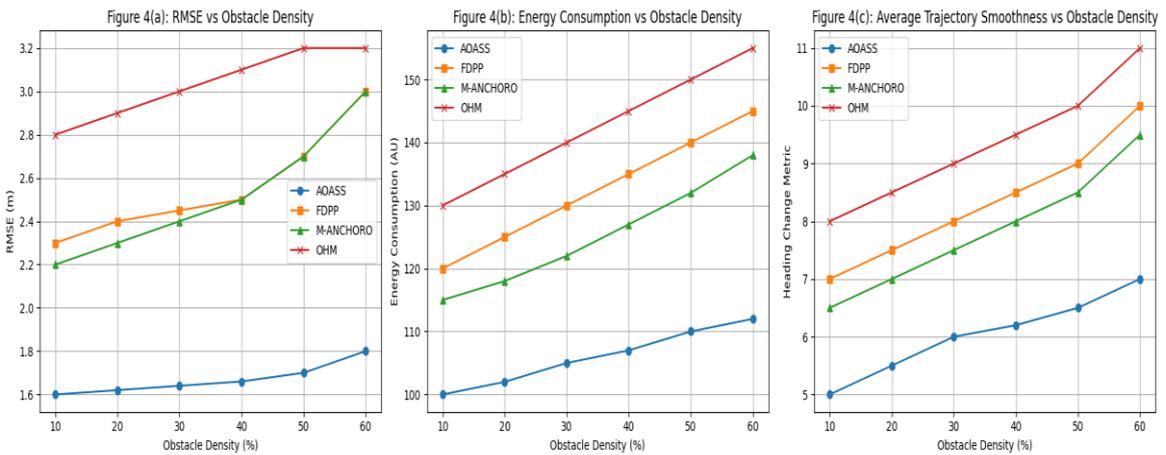

Figure 4. Robustness to Obstacle Density

### 4.2.5. Scalability Analysis

To evaluate the scalability of the AOASS framework, simulations were conducted with increased node densities ranging from 600 to 1000 nodes, as well as expanded sensing fields up to 200 × 200 m². Figure 3(a) illustrates the impact of network size on localization accuracy. AOASS maintains a relatively low RMSE (≈ 1.5–1.7 m) even at the highest node density, outperforming FDPP (≈ 2.3–2.5 m), M-ANCHORO (≈ 2.1–2.4 m), and OHM (≈ 2.7–3.0 m). Energy consumption and trajectory efficiency were also analysed (Figures 3(b) and 3(c)). Despite the larger network size, AOASS exhibits moderate increases in energy use and path length, thanks to the dynamic path adaptation of the Adaptive Square-Spiral Trajectory and the reinforcement-learning-based motion controller. The results confirm that AOASS scales efficiently while preserving high localization accuracy and coverage in denser or larger networks.

### 4.2.6. Robustness to Obstacle Density

The robustness of AOASS against high obstacle density was investigated by increasing obstacle ratios to 50% and 60%, representing highly cluttered environments. Figure 4(a) shows that AOASS maintains superior localization accuracy (≈ 1.6–1.8 m), whereas FDPP and M-ANCHORO experience significant degradation (≈ 2.5–3.0 m). OHM performance deteriorates further (≈ 3.2 m). Energy efficiency under dense obstacles was evaluated in Figure 4(b), highlighting that AOASS sustains lower energy consumption (≈ 20–25% savings over FDPP) due to the fuzzy-based ORCA module, which efficiently avoids collisions without unnecessary detours. Trajectory analysis (Figure 4(c)) confirms that AOASS generates smooth, collision-free paths with minimal abrupt heading changes, even in highly obstructed fields.

## 5. CONCLUSIONS

This study introduced AOASS (Adaptive Obstacle-Aware Square Spiral), a novel and intelligent localization framework designed for Wireless Sensor Networks (WSNs) operating in environments with static and dynamic obstacles. The proposed system integrates a square spiral trajectory optimized for full coverage with an adaptive obstacle-awareness mechanism, enabling the mobile anchor to efficiently navigate the sensing field while maintaining reliable connectivity with sensor nodes. By combining the OLSTM-DV-Hop localization algorithm—an enhanced hybrid approach that fuses the classical DV-Hop with Long Short-Term Memory (LSTM) prediction—the proposed model significantly improves hop-distance estimation accuracy and minimizes multihop propagation errors. Furthermore, the TD3-LSTM-based trajectory planner dynamically optimizes the anchor's movement path according to real-time environmental feedback, while the Kalman-based predictive tracking and FLC-ORCA reactive layer ensure safe and energy-efficient navigation around obstacles.

Simulation results confirm that AOASS consistently outperforms conventional localization schemes in terms of localization accuracy, energy efficiency, coverage ratio, and trajectory optimization, particularly in static environments with dense and irregularly distributed obstacles. These improvements are attributed to the model's adaptive learning capability, which allows it to intelligently balance exploration, energy usage, and obstacle avoidance without sacrificing localization precision. Overall, AOASS offers a scalable, AI-driven localization solution suitable for real-world IoT and WSN applications. Its hybrid integration of deep learning, reinforcement learning, and intelligent control mechanisms provides a robust foundation for autonomous and context-aware network operations, paving the way for next-generation smart sensing and adaptive localization systems. Future work will focus on extending the proposed framework to handle dynamic obstacle scenarios, multi-anchor coordination, and real-time adaptability, further enhancing its applicability to complex and evolving IoT environments.

CONFLICTS OF INTEREST

The author declares no conflict of interest.

International Journal of Computer Networks & Communications (IJCNC) Vol.18, No.1, January 2026

**AUTHOR**

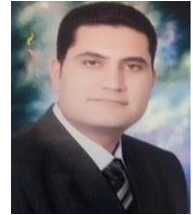

**Abdelhady Naguib** received his M.S. and Ph.D. in Systems and Computers Engineering from Al-Azhar University in 2008 and 2013 respectively. He is currently an associate professor at, Faculty of Engineering at Al-Azhar University, Cairo, Egypt. In addition, he is working as assistant professor at department of Computer Science, Jouf University, Saudi Arabia.

He has a research and teaching experience for many years. He is also acting as a reviewer of well-reputed journals like Springer Nature and Inderscience Online. His research area is in the field of Mobile Communications more specific in the area of mobile ad hoc and wireless sensor networks. He has contributed to this area by many papers and one book, "Extending NS-2 for Simulating a Localization Algorithm," 1st. ed., LAP LAMBERT Academic Publishing, April 2017, ISBN-13 978-3-330-07674-7. Dr. Abdelhady Naguib is a member at Egyptian Syndicate for Engineers and has a membership of Cisco academy (as an instructor).